\documentclass[sigconf]{acmart}  \renewcommand\footnotetextcopyrightpermission[1]{} \acmConference{}{}{}   
 \renewcommand\footnotetextcopyrightpermission[1]{} 
\usepackage{amsmath}
\usepackage{pgfplots}
\usepackage{float}
\usepackage{multirow}

\begin{document}

\title{Intelligent Elastic Feature Fading: Enabling Model Retrain-Free Feature Efficiency Rollouts at Scale}


\author{Jieming Di}
\affiliation{
  \institution{Meta}\country{USA}
}
\email{jiemingdi@meta.com}

\author{Xiaoyu Chen}
\affiliation{
  \institution{Meta}\country{USA}
}
\email{xochen@meta.com}

\author{Ying She}
\affiliation{
  \institution{Meta}\country{USA}
}
\email{yingshe@meta.com}

\author{Siyu Wang}
\affiliation{
  \institution{Meta}\country{USA}
}
\email{siyuw@meta.com}

\author{Lizzie Liu}
\affiliation{
  \institution{Meta}\country{USA}
}
\email{lizzieliu@meta.com}

\author{Fenggang Wu}
\affiliation{
  \institution{Meta}\country{USA}
}
\email{fwu@meta.com}

\author{Jiaoying Mu}
\affiliation{
  \institution{Meta}\country{USA}
}
\email{joymu@meta.com}

\author{Tony Tsui}
\affiliation{
  \institution{Meta}\country{USA}
}
\email{tonytsui@meta.com}

\author{Amr Elroumy}
\affiliation{
  \institution{Meta}\country{USA}
}
\email{aae@meta.com}

\author{Hsing Tang}
\affiliation{
  \institution{Meta}\country{USA}
}
\email{hsingtang@meta.com}

\author{Zewei Jiang}
\affiliation{
  \institution{Meta}\country{USA}
}
\email{zeweijiang@meta.com}

\author{Qiao Yang}
\affiliation{
  \institution{Meta}\country{USA}
}
\email{qyang@meta.com}

\author{Lin Qi}
\affiliation{
  \institution{Meta}\country{USA}
}
\email{linqi@meta.com}

\author{Haibo Lin}
\affiliation{
  \institution{Meta}\country{USA}
}
\email{haibolin@meta.com}

\author{Weifeng Cui}
\affiliation{
  \institution{Meta}\country{USA}
}
\email{weifengc@meta.com}

\author{Daniel Li}
\affiliation{
  \institution{Meta}\country{USA}
}
\email{lidanielli@meta.com}

\author{Kapil Gupta}
\affiliation{
  \institution{Meta}\country{USA}
}
\email{kapilg@meta.com}

\author{Shivendra Pratap Singh}
\affiliation{
  \institution{Meta}\country{USA}
}
\email{shiven@meta.com}

\author{Jie Zheng}
\affiliation{
  \institution{Meta}\country{USA}
}
\email{jie2@meta.com}

\author{Arnold Overwijk}
\affiliation{
  \institution{Meta}\country{USA}
}
\email{arnoldov@meta.com}

\author{Ling Leng}
\affiliation{
  \institution{Meta}\country{USA}
}
\email{lleng@meta.com}

\author{Sri Reddy}
\affiliation{
  \institution{Meta}\country{USA}
}
\email{sriharir@meta.com}

\author{Robert Malkin}
\affiliation{
  \institution{Meta}\country{USA}
}
\email{rgmalkin@meta.com}

\author{Rocky Liu}
\affiliation{
  \institution{Meta}\country{USA}
}
\email{rockyliu4@meta.com}

\begin{abstract}
Large-scale ranking systems depend on thousands of features derived from user behavior across multiple time horizons. Typically requires model retraining---resulting in long iteration cycles (3--6 months), substantial GPU resource consumption, and limited rollout throughput.

We introduce Intelligent Elastic Feature Fading (IEFF), a production infrastructure system that enables retrain-free feature efficiency rollouts by elastically controlling feature coverage and distribution at serving time. IEFF supports incremental feature coverage adjustments while models adapt through recurring training, eliminating dependencies on explicit retraining cycles. The system incorporates strict safety guardrails, reversibility mechanisms, and comprehensive monitoring to ensure stability at scale.

Across multiple production use cases, IEFF accelerates efficiency-related rollouts by 5$\times$, eliminates retraining-related GPU overhead, and enables faster capacity recycling. Extensive offline and online experiments demonstrate that gradual feature fading prevents 50--55\% of online performance degradation compared to abrupt feature removal, while maintaining stable model behavior. These results establish elastic, system-level feature fading as a practical and scalable approach for managing feature efficiency in modern industrial ranking systems.
\end{abstract}

\keywords{feature management, ranking systems, feature efficiency, production ML systems, model serving}

\renewcommand{\shortauthors}{} \renewcommand{\headrulewidth}{0pt} \fancyhead{}
\maketitle

\section{Introduction}

Large-scale ranking systems rely on thousands of features derived from user behavior across multiple time horizons. Prior work has shown that feature design and management play a role in the effectiveness of industrial ranking and ads systems~\cite{he2014predictingclicks,naumov2019dlrm}. While these features are essential for model quality, they also introduce infrastructure cost in data storage, training, and online serving. As a result, feature efficiency techniques---such as feature deprecation, feature space cleanup, and precision migration---have become increasingly important for operating ranking systems at scale.

Despite their importance, the rollout of feature efficiency techniques has become a major operational bottleneck. Many efficiency changes require modifying feature coverage or distribution, which is commonly gated by model retraining to avoid performance regressions. In practice, retraining must align with predefined model cycle schedules, resulting in long iteration cycles that often span several months. Moreover, retraining incurs substantial GPU resource consumption, significantly increasing operational overhead and delaying capacity recycling. Prior work has highlighted the high computational cost and operational burden of frequent retraining in large-scale machine learning systems~\cite{mahadevan2023cost,sculley2015techdebt}.

Prior studies on model maintenance show that abrupt changes in data or feature distributions---often referred to as concept or covariate drift---can cause severe model instability and performance degradation if not properly managed~\cite{gama2014survey,mallick2022drift,dong2024drift}. A common mitigation strategy is to retrain models upon detecting drift; however, this approach further exacerbates computational cost and rollout latency in production systems.

At the same time, modern ranking models exhibit strong memorization capacity and are trained continuously on fresh data through recurring training pipelines~\cite{baylor2017tfx}. When feature coverage or distribution is adjusted gradually, models can adapt to these changes without requiring explicit retraining cycles. This observation suggests that retraining is not inherently necessary for many feature efficiency rollouts, provided that changes are applied in a controlled and incremental manner.

Based on this insight, we present Intelligent Elastic Feature Fading (IEFF), a system-level infrastructure capability that enables model retrain-free feature efficiency rollouts. IEFF introduces elastic control over feature coverage and distribution at the serving stage, allowing changes to be deployed incrementally while preserving training--serving consistency. The system incorporates strict safety guardrails, reversibility, and continuous monitoring to ensure stable operation at production scale, consistent with best practices in modern ML infrastructure~\cite{polyzotis2019datavalidation,caveness2020tfdv}.

IEFF accelerates feature efficiency rollouts by 5$\times$, eliminates retraining-related GPU overhead, and enables faster capacity recycling. Extensive offline and online experiments show that gradual feature fading prevents up to 50--55\% of online performance loss compared to feature zero-out, while maintaining stable model behavior under controlled rollouts validated through online experimentation~\cite{kohavi2013experiments}.

Our contributions are:
\begin{itemize}
  \item We introduce IEFF, a production infrastructure system that enables retrain-free feature efficiency rollouts by elastically controlling feature coverage at serving time.
  \item We demonstrate that IEFF significantly improves rollout velocity while reducing GPU and operational overhead in large-scale ranking systems.
  \item We provide empirical evidence showing that gradual feature fading substantially improves stability compared to abrupt feature zero-out.
\end{itemize}

\section{Problem \& Key Insight}
\subsection{Problem: Abrupt Feature Changes and Retraining Dependency}

In large-scale ranking systems, feature rollouts are often tightly coupled with model configuration updates and explicit retraining, for example when deprecating features or replacing existing features. Such coupling is commonly adopted to prevent performance regressions in production systems, but it introduces long iteration cycles, high GPU cost, and limited rollout throughput~\cite{mahadevan2023cost,sculley2015techdebt}.

One key reason for this dependency is that abrupt feature changes — such as setting feature coverage directly from 100\% to 0\% (feature zero-out) — can cause severe model instability. Prior work on concept and covariate drift has shown that sudden distribution shifts can significantly degrade model performance if not properly managed~\cite{gama2014survey,mallick2022drift,dong2024drift}. 

\subsection{Observation: Gradual Feature Changes Enable Model Adaptation}

Despite their sensitivity to abrupt feature changes, modern ranking models are continuously retrained on freshly logged data through recurring training pipelines~\cite{baylor2017tfx}. This creates an important observation: when feature coverage or distribution is adjusted gradually, models can often adapt through recurring training without requiring explicit retraining cycles tied to feature rollouts.

Intuitively, incremental feature changes bound the magnitude of distribution shift encountered at each training step, allowing model parameters to update smoothly over time. Because the training pipeline continuously incorporates newly logged data, the updated feature distribution is gradually reflected in the training dataset, enabling the model to adjust its parameters across successive training iterations. In effect, the model can track the evolving feature distribution without experiencing large optimization shocks.

As a result, gradual feature changes typically avoid the sharp increases in normalized entropy (NE) observed in abrupt feature zero-out scenarios, where large instantaneous distribution shifts are introduced and must be corrected through explicit retraining, often at substantial computational cost~\cite{mahadevan2023cost}.

\subsection{Key Insight}

Based on the observations above, we identify the following key insight:

\begin{quote}
\emph{Explicit model retraining is not inherently required for many feature rollouts, provided that feature coverage or distribution changes are applied incrementally and models are allowed to adapt through recurring training.}
\end{quote}

This insight suggests that the retraining dependency commonly observed in feature rollouts is largely a consequence of how feature changes are applied, rather than an intrinsic requirement of the models themselves. Modern ML infrastructure increasingly emphasizes continuous training pipelines, validation, and monitoring to maintain training--serving consistency at scale~\cite{baylor2017tfx,polyzotis2019datavalidation,caveness2020tfdv}, which makes it feasible to operationalize this insight at the system level.

To realize this insight in production systems, however, requires an infrastructure mechanism that can safely, incrementally, and reversibly control feature coverage and distribution over time while preserving training--serving consistency.

In the next section, we introduce Intelligent Elastic Feature Fading (IEFF), a production infrastructure system designed to realize this insight by enabling controlled, incremental feature changes at serving time.

\begin{figure*}[t]
  \centering
  \includegraphics[width=0.95\textwidth]{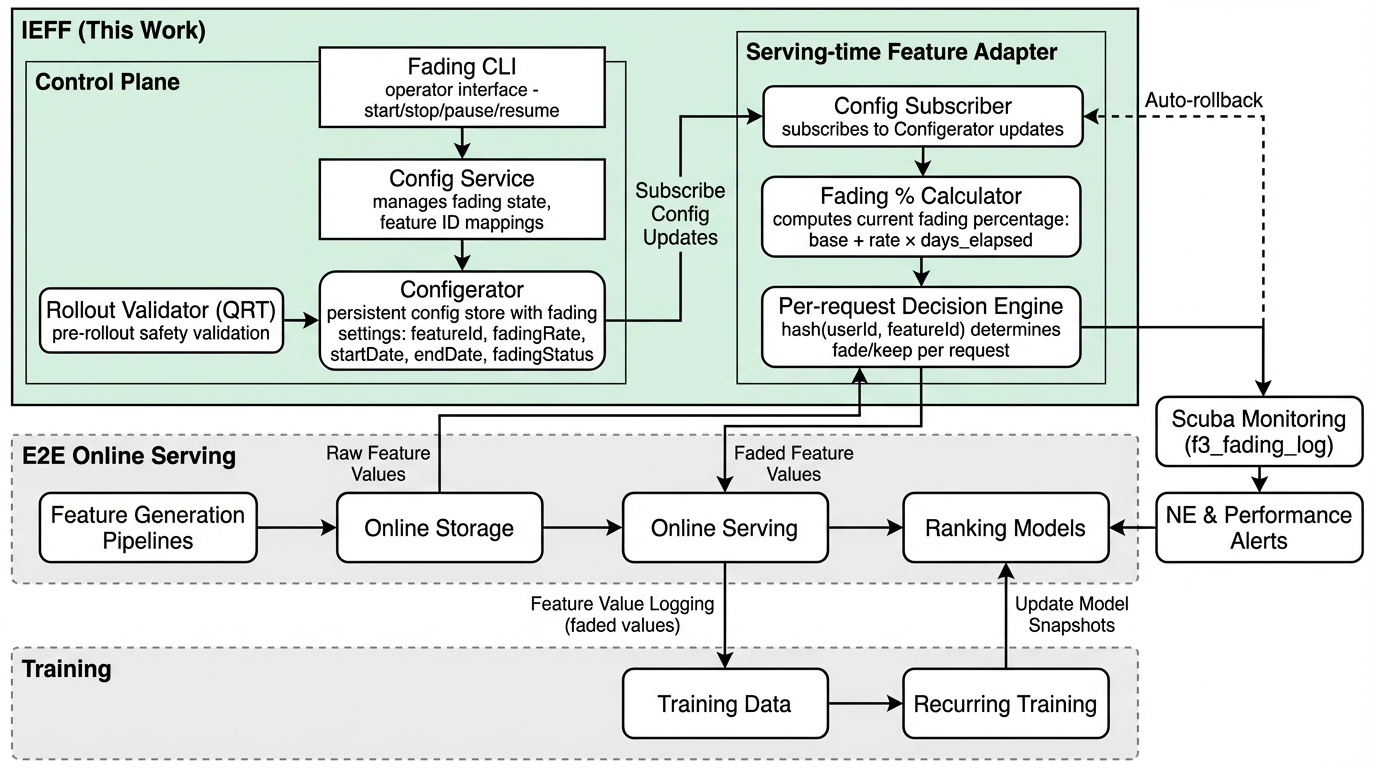}
  \caption{IEFF system architecture. IEFF introduces a centralized control plane and a serving-time feature adapter that incrementally controls feature coverage during online serving. The adjusted feature values are used consistently for inference and logged for recurring training, enabling retrain-free and reversible feature rollouts.}
  \label{fig:ieff_arch}
\end{figure*}

\section{IEFF System Design}
\subsection{Design Goals}

IEFF is a system-level infrastructure capability designed to decouple feature rollouts from explicit model retraining in large-scale ranking systems. Similar to prior production ML platforms, IEFF emphasizes modularity, safety, and scalability as first-class design considerations~\cite{baylor2017tfx,sculley2015techdebt}. Its design is guided by the following goals.

\textbf{Retrain-free rollout.}  
IEFF enables feature rollouts without blocking on explicit retraining cycles by supporting incremental changes to feature coverage or distribution, reducing dependency on costly retraining workflows~\cite{mahadevan2023cost}.

\textbf{Safety and stability.}  
The system enforces controlled rollout behaviors to avoid abrupt distribution shifts that could lead to sharp increases in normalized entropy (NE) or performance regressions, a known challenge in production ML systems under distribution drift~\cite{gama2014survey,mallick2022drift}.

\textbf{Reversibility.}  
All rollouts must be reversible, allowing rapid rollback to previous configurations without requiring additional retraining, consistent with best practices for managing risk in large-scale ML systems~\cite{sculley2015techdebt}.

\textbf{Model-agnostic operation.}  
IEFF operates transparently across different model architectures and training pipelines, without relying on model-specific assumptions, enabling reuse across heterogeneous ranking models~\cite{abadi2016tensorflow}.

\textbf{Production scalability.}  
The system is designed to scale to production environments serving billions of requests per day with minimal latency and operational overhead, a requirement common to industrial ranking and ads platforms~\cite{he2014predictingclicks,naumov2019dlrm}.

\subsection{Architecture Overview}

Figure~\ref{fig:ieff_arch} provides a high-level overview of the IEFF architecture. IEFF introduces a system-level control plane and a serving-time feature adapter to decouple feature rollouts from explicit model retraining, while leaving existing feature generation pipelines unchanged. This design aligns with modern end-to-end ML platforms that separate control logic from data paths~\cite{baylor2017tfx}.

Feature generation remains identical to existing production setups and produces raw feature values with their original coverage and distribution. IEFF does not modify offline or online feature pipelines, which allows the system to be deployed incrementally without disrupting existing feature authoring or data processing workflows, a common requirement in large-scale ML infrastructures~\cite{orr2021featurestores}.

IEFF introduces a centralized control plane that manages rollout policies, including fading configurations, rollout state, and safety constraints. These policies are enforced by a serving-time feature adapter, which adjusts the effective feature coverage or distribution according to the configured rollout schedule. Importantly, this adjustment occurs at serving time, allowing feature changes to be applied incrementally without triggering explicit model retraining.

IEFF preserves serving--training consistency by design: the adjusted feature values are used for both online inference and downstream training, ensuring the model always trains on the same inputs it served. This avoids training--serving skew and aligns with established best practices in continuous ML pipelines~\cite{polyzotis2019datavalidation,caveness2020tfdv}.

Models perform inference using the adjusted feature values and continuously update through recurring training on newly logged data. Monitoring and rollback mechanisms operate alongside the control plane to track key system metrics, such as normalized entropy (NE) and online performance, and to revert rollouts when safety thresholds are violated.

\subsection{Fading Control Mechanism}

IEFF realizes incremental feature rollouts through a fading control mechanism that gradually adjusts feature coverage or distribution over time. Rather than applying abrupt changes, fading decomposes a rollout into a sequence of small, controlled updates that are enforced at serving time, mitigating the risks associated with sudden distribution shifts~\cite{gama2014survey,mallick2022drift}.

At a high level, fading is parameterized by a rollout schedule defined by a start time and a fading rate. Once configured, the rollout proceeds automatically without requiring additional operator intervention. The fading rate determines how quickly the target feature configuration changes over time.

In practice, the fading rate is determined based on results from pre-rollout validation via Q/R testing (QRT)---an internal A/B testing framework---which evaluates the impact of gradual feature changes under controlled conditions. QRT allows operators to assess model sensitivity to incremental feature changes and select a safe rollout rate before deployment. In practice, we have validated fading rates ranging from 1\% to 10\% per day across different feature types and models. Conservative rates (1--2\% per day) are used for production rollouts to ensure that distribution shifts remain incremental and within safe bounds, while higher rates (5--10\%) are applied in boundary experiments and time-sensitive scenarios such as privacy-related deprecations.

IEFF distinguishes between controlling feature coverage and feature distribution. Coverage control determines whether a feature is present for a given request, while distribution control modifies the effective values of a feature without fully removing it. Both forms of control are applied at serving time and can be combined to support a wide range of rollout scenarios, such as feature deprecation, feature migration, and feature space cleanup.

Fading is enforced at the serving layer rather than within feature generation pipelines. This design decouples feature rollouts from upstream data processing and avoids the need to re-materialize features or trigger explicit model retraining, which are known sources of operational overhead in production ML systems~\cite{mahadevan2023cost,sculley2015techdebt}. Because the same effective feature values are used for inference and logged for downstream training, fading integrates naturally with continuous training pipelines and preserves serving--training consistency.

The fading mechanism is intentionally designed to be model-agnostic. It does not rely on assumptions about model architecture or feature semantics, allowing the same rollout mechanism to be reused across heterogeneous models and feature types.

\subsection{Safety, Governance, and Rollback}

Because IEFF operates at the system level and directly affects feature values used by production models, safety and governance are first-class design considerations, consistent with prior observations on managing risk and technical debt in ML systems~\cite{sculley2015techdebt}.

Before any production rollout, IEFF requires feature changes to be validated through fading QRT, which evaluates the impact of gradual feature changes under controlled conditions. QRT results are used to determine safe fading rates and to verify that the rollout does not introduce unacceptable distribution shifts or system instability. This pre-rollout validation serves as a safety gate and complements established practices in controlled online experimentation~\cite{kohavi2013experiments}.

In production, IEFF restricts fading to explicitly designated features and executes rollouts under predefined policies that constrain rollout scope, fading rate, and duration. These constraints ensure that all feature changes remain incremental and bounded.

During rollouts, IEFF continuously monitors key system metrics, including normalized entropy (NE) and business-facing performance indicators, enabling early detection of abnormal behavior through validation and monitoring mechanisms~\cite{polyzotis2019datavalidation}. When monitored metrics exceed predefined safety thresholds, rollouts can be automatically paused or rolled back. Operators can also trigger manual rollback when abnormal patterns are observed.

IEFF supports rapid rollback by design. Fading configurations can be reverted at any point during a rollout, immediately restoring original feature coverage or distribution without requiring model retraining or pipeline changes.

Together, these mechanisms allow IEFF to enable retrain-free feature rollouts while maintaining system stability and operational control.

\subsection{System Overhead and Efficiency}

IEFF introduces a lightweight serving-time feature adapter that applies deterministic fading decisions without modifying upstream feature generation pipelines. The adapter operates on cached rollout configurations and performs simple gating or value adjustment, incurring negligible computational overhead.

In production, we observe no measurable increase in serving latency attributable to IEFF. The additional logic executes within the existing feature processing path and does not introduce extra network calls or synchronization barriers. Control-plane updates are infrequent and propagate asynchronously, ensuring that rollout configuration changes do not impact request critical paths.

IEFF also preserves training--serving consistency by logging the post-fading feature values observed during inference, ensuring the model's continuous training consumes the same values used for serving. This avoids additional storage overhead or duplicated logging. Overall, IEFF enables retrain-free rollouts with minimal runtime and operational cost.

\section{Workflow and Use Cases}

IEFF is designed as a general-purpose infrastructure capability for managing feature rollouts in large-scale ranking systems. Similar to prior production ML platforms, IEFF emphasizes reuse across different rollout scenarios while preserving safety and scalability~\cite{baylor2017tfx}. In this section, we describe several representative use cases that illustrate how IEFF can be applied in production workflows.

\subsection{Feature Deprecation}

Feature deprecation is a common operation in production systems, driven by factors such as feature obsolescence, redundancy, or policy requirements. Traditional deprecation workflows often rely on abrupt feature zero-out followed by explicit model retraining, which can introduce sudden distribution shifts and lead to performance degradation~\cite{gama2014survey}.

With IEFF, feature deprecation can be implemented through gradual fading of feature coverage or distribution at serving time. By reducing feature influence incrementally, models can adapt through recurring training without requiring explicit retraining cycles. This approach mitigates sharp increases in normalized entropy (NE) and reduces the risk of instability during deprecation. Rollout safety can be validated using controlled online experimentation prior to full deployment~\cite{kohavi2013experiments}.

\subsection{Feature Migration and Efficiency Rollouts}

Feature migration and efficiency-driven rollouts, such as replacing features with more compact representations or optimizing feature storage, frequently require coordinated changes across models and pipelines. In practice, these rollouts are often delayed by retraining schedules and operational overhead.

IEFF enables such rollouts to be performed incrementally by fading out legacy features while fading in new or optimized feature representations. Because feature changes are applied at serving time and logged consistently for training, models can transition smoothly between feature versions without blocking on retraining, significantly accelerating rollout timelines.

\subsection{Privacy and Emergency Rollouts}

In certain scenarios, such as privacy-related feature removal or emergency mitigations, feature changes must be applied rapidly while minimizing negative impact on model performance. Abrupt feature removal in these cases can lead to substantial performance regressions due to sudden distribution shifts.

IEFF provides a controlled mechanism for handling such scenarios by allowing feature influence to be reduced progressively under strict safety constraints. Combined with pre-rollout validation and continuous monitoring, IEFF enables faster response times while preserving system stability, consistent with best practices in production ML validation pipelines~\cite{polyzotis2019datavalidation}.

\subsection{Deployment Summary}

Table~\ref{tab:deployment_summary} summarizes IEFF deployment across three production rollout phases. The system faded 275 features across 14 batches, with rates refined from 10\%/day (early validation) to 1--5\%/day (large-scale rollouts). Each fading batch avoids approximately 10 full model retrains on average, as models adapt through recurring training. In total, IEFF eliminated an estimated 140 retraining iterations and reduced annual infrastructure costs by approximately 15\%.

\begin{table}[t]
\centering
\small
\caption{Cumulative IEFF deployment metrics by year.}
\label{tab:deployment_summary}
\begin{tabular}{lrrrl}
\toprule
\textbf{Year} & \textbf{\#Feat.} & \textbf{Rate} & \textbf{Savings} & \textbf{Retrains} \\
 & & \textbf{(\%/day)} & \textbf{(\%)} & \textbf{Avoided} \\
\midrule
2024 & 3 & 10 & 0.3\% & $\sim$20 \\
2025 & 135 & 0--10 & 6\% & $\sim$70 \\
2026 & 137 & 2--5 & 9\% & $\sim$50 \\
\midrule
\textbf{Total} & \textbf{275} & 0--10 & \textbf{$\sim$15\%} & $\sim$\textbf{140} \\
\bottomrule
\end{tabular}
\end{table}

\section{Evaluation}

We evaluate IEFF on an ads ranking platform to answer three questions:
(1) whether incremental feature fading improves stability compared to abrupt zero-out,
(2) whether IEFF enables retrain-free rollouts without degrading online performance, and
(3) how IEFF affects rollout efficiency and operational cost.

\subsection{Experimental Setup}

Experiments are conducted on production ranking models. We evaluate IEFF across multiple model types, including click-through rate (CTR) and conversion rate (CVR) prediction models, and across different feature types, including sparse ID-based features and embedding-based features. This diversity allows us to assess the generalizability of IEFF across heterogeneous ranking components.

For offline evaluation, we simulate feature fading during recurring training using predefined fading schedules with daily fading rates ranging from 1\% to 10\%. For online evaluation, we conduct controlled A/B experiments (QRT) comparing IEFF-based fading against zero-out. Across all experiments, we monitor normalized entropy (NE) as a stability metric and online performance as the primary business metric.

\subsection{Stability and Online Performance}

\begin{figure}[t]
  \centering
  \includegraphics[width=0.9\linewidth]{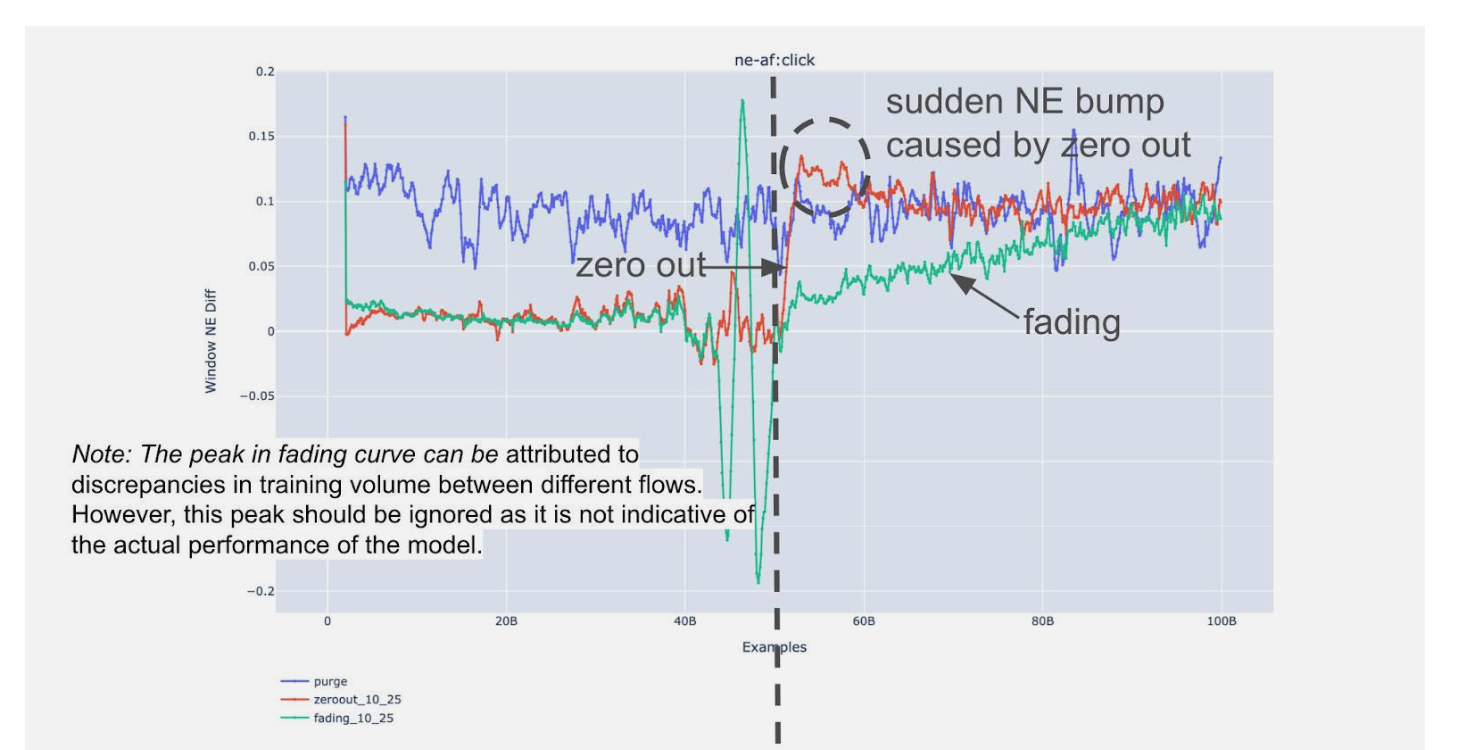}
  \caption{Offline recurring training results. Gradual feature fading reduces peak and cumulative normalized entropy (NE) increase compared to abrupt feature zero-out.}
  \label{fig:fading_results}
\end{figure}

Figure~\ref{fig:fading_results} and Table~\ref{tab:online_performance} summarize the offline and online results, respectively.

Figure~\ref{fig:fading_results} shows that, under recurring training, gradual feature fading substantially reduces peak and cumulative online performance degradation compared to abrupt zero-out.

Table~\ref{tab:online_performance} reports offline results across different feature types and models. All configurations consistently show $\sim$50\% reduction in daily performance loss, regardless of feature type or model.

We further validate these findings with online experiments. During the rollout window, zeroing out the top 50 sparse features results in a 0.83\% online performance regression, whereas gradual fading leads to a 0.37\% regression, reducing the rollout-induced performance loss by approximately 55\%.

Additionally, abrupt zero-out caused a transient performance spike requiring 7--10 billion additional training samples to recover, whereas gradual fading converged smoothly without recovery delays. In practice, abrupt feature removal has been linked to multiple production incidents caused by sudden performance spikes; IEFF's gradual mechanism serves as an additional safeguard against such instability.

While the above results demonstrate that IEFF improves stability and online performance on average, safe deployment in production also requires effective guardrails to detect and mitigate abnormal behaviors during rollouts.

\begin{table}[t]
  \centering
  \caption{Offline daily absolute increase in normalized entropy (NE)
  during the fading window (coverage from 100\% to 0\%).
  All configurations show $\sim$50\% reduction in daily NE increase
  under fading compared to zero-out.}
  \label{tab:online_performance}
  \small
  \begin{tabular}{lcc}
    \toprule
    \textbf{Configuration} & \textbf{Zero-out} & \textbf{Fading} \\
    \midrule
    Top 50 sparse (CTR) & 0.10\%/day & 0.05\%/day \\
    Top 30 sparse (CTR) & 0.04\%/day & 0.02\%/day \\
    Top 2 embed. (CVR)  & 0.15\%/day & 0.075\%/day \\
    \bottomrule
  \end{tabular}
\end{table}

\subsection{Phase-wise Online Performance Comparison}

To further understand how gradual fading differs from abrupt zero-out during a rollout,
we analyze a decreasing-coverage scenario in which feature coverage is reduced from 90\% to 0\%
over multiple days.
Table~\ref{tab:zeroout_vs_fading_delta} presents a phase-wise comparison of online performance,
where all values are normalized to the gradual fading baseline (100\%).

Across all rollout phases, gradual fading consistently maintains stable online performance,
remaining close to the normalized baseline throughout the entire rollout.
This indicates that models are able to continuously adapt to incremental feature changes
through recurring training without experiencing abrupt performance degradation.

In contrast, zero-out exhibits systematic performance loss at every phase of the rollout.
The degradation is most pronounced during the mid-coverage phase (70\%--40\%),
where zero-out underperforms fading by approximately 0.6\%.
This phase corresponds to the regime in which the feature is still partially relied upon
by the model, but its sudden removal introduces a large and poorly conditioned
distribution shift.

Notably, the performance gap between zero-out and fading narrows again in the late and final
phases of the rollout, after feature coverage has already been substantially reduced.
This behavior suggests that once the model has fully adapted to the absence of the feature,
the impact of further removal becomes less severe.
However, the transient degradation observed in earlier phases can accumulate
into meaningful online impact during production rollouts.

These results highlight that the primary benefit of IEFF lies not only in the final steady state,
but also in stabilizing the intermediate rollout trajectory.
By smoothing distribution shifts during the most sensitive phases,
gradual fading reduces transient performance loss and lowers the operational risk
associated with large-scale feature rollouts.

Overall, this phase-wise analysis complements the aggregate QRT results in Section~5.2
and demonstrates that IEFF is particularly effective in mitigating instability
during intermediate rollout stages, which are often the most challenging to manage
in production environments.

\begin{table}[t]
  \centering
  \caption{Online performance comparison between abrupt zero-out and gradual fading,
  derived from the decreasing-coverage rollout (Figure~\ref{fig:fading_results}).
  Performance is normalized to fading (100\%).}
  \label{tab:zeroout_vs_fading_delta}
  \small
  \begin{tabular}{lcccc}
    \toprule
    \textbf{Rollout Phase} & \textbf{Coverage} & \textbf{Zero-out} & \textbf{Fading} & \textbf{Delta} \\
    \midrule
    Early  & 90\%--70\% & $\sim$99.8\% & 100\% & $-0.2\%$ \\
    Mid    & 70\%--40\% & $\sim$99.4\% & 100\% & $-0.6\%$ \\
    Late   & 40\%--10\% & $\sim$99.6\% & 100\% & $-0.4\%$ \\
    Final  & 10\%--0\%  & $\sim$99.7\% & 100\% & $-0.3\%$ \\
    \bottomrule
  \end{tabular}
\end{table}

\subsection{Rollout Efficiency}

We further evaluate the operational impact of IEFF on rollout efficiency. Traditional retraining-gated rollouts require waiting for scheduled retraining cycles and incur significant GPU overhead. In contrast, IEFF enables feature rollouts to be executed immediately through serving-time configuration changes.

In production, IEFF accelerates feature efficiency rollouts by approximately 5$\times$ and eliminates retraining-related GPU cost during rollouts, enabling faster capacity recycling and higher rollout throughput.

\subsection{Summary}

Overall, IEFF enables retrain-free feature rollouts with improved stability, reduced online performance regression, and substantially lower operational cost compared to retraining-based approaches.

\section{Discussion \& Lessons Learned}

Through the design and deployment of IEFF, we summarize several practical lessons for enabling retrain-free feature rollouts in large-scale ranking systems.

\paragraph{Incremental change and scope.}
We find that the rate of feature coverage or distribution change often matters more than the final target state. Applying changes incrementally allows models to adapt through recurring training and significantly reduces instability. IEFF is most effective for feature deprecations, migrations, and efficiency-driven rollouts where changes can be decomposed into bounded steps.

\paragraph{Serving-time control.}
Applying fading at the serving layer proved essential for decoupling feature rollouts from upstream pipelines and retraining schedules. This design minimizes cross-team coordination overhead and enables rapid rollback through configuration changes.

\paragraph{Training--serving consistency.}
Retrain-free rollouts critically rely on preserving training--serving consistency. IEFF ensures that the same effective feature values used for inference are logged and consumed by recurring training, allowing models to adapt safely over time.

\paragraph{Safety guardrails.}
Incremental rollouts require strict safety controls, including pre-rollout validation, bounded fading rates, continuous monitoring, and fast rollback. In practice, these guardrails often determine whether retrain-free rollouts are feasible for a given feature.

\paragraph{Summary.}
Overall, our experience shows that retraining dependencies in feature rollouts are often a consequence of how changes are applied rather than an inherent limitation of models.

\section{Related Work}

\paragraph{Production ML Systems and Feature Management.}
Large-scale machine learning systems rely on complex feature pipelines to support training and online serving at scale. Prior work has emphasized modular pipelines, validation, and monitoring to ensure reliability and training--serving consistency in production environments~\cite{baylor2017tfx,sculley2015techdebt,orr2021featurestores}. Feature stores have emerged as a key infrastructure abstraction for managing feature computation, sharing, and consistency across training and serving~\cite{orr2021featurestores}. IEFF builds on these system design principles but focuses specifically on enabling incremental, serving-time feature rollouts, a capability not directly addressed by existing feature management frameworks.

\paragraph{Feature Selection and Efficiency in Ranking Systems.}
Feature selection and efficiency are important for managing complexity in large-scale ranking and recommendation systems. Classical approaches include filter, wrapper, and embedded methods for identifying informative features~\cite{guyon2003introduction}. In deep recommendation models, structured pruning and feature interaction selection have been explored to reduce model size while preserving quality~\cite{deng2021deeplight,wang2022dcnv2}. These methods typically require retraining after feature changes. IEFF addresses a complementary problem: enabling safe, retrain-free rollouts of feature efficiency decisions that have already been made.

\paragraph{Retraining Cost and Model Lifecycle Management.}
Several studies have examined the computational and operational cost of retraining models in large-scale systems, identifying retraining as a major bottleneck for rapid iteration~\cite{mahadevan2023cost}. Model lifecycle and deployment systems have been proposed to organize retraining and release workflows~\cite{sun2020gallery}. Continual and incremental learning methods aim to update models without full retraining~\cite{parisi2019continual}, but these focus on model-level adaptation rather than system-level feature rollout control. In contrast, IEFF reduces reliance on explicit retraining by decoupling feature rollouts from retraining cycles through serving-time control.

\paragraph{Distribution Shift and Stability.}
Concept drift and covariate drift are well-known causes of model degradation under changing data distributions~\cite{gama2014survey}. Recent system-oriented work has explored detecting and mitigating drift in large-scale ML pipelines, often relying on retraining as the primary response mechanism~\cite{mallick2022drift,dong2024drift}. Gradual deployment strategies, such as canary releases and staged rollouts, are widely adopted in software engineering~\cite{schermann2018continuous} but have received limited attention in the context of ML feature management. IEFF complements this line of work by limiting the magnitude of distribution shift introduced at each step through incremental feature changes, enabling models to adapt via recurring training without triggering explicit retraining.

\paragraph{Online Experimentation and Safe Deployment.}
Controlled online experimentation is a cornerstone of safe deployment in large-scale systems~\cite{kohavi2013experiments}. Prior work has established frameworks for A/B testing, metric monitoring, and automated rollback in production environments~\cite{tang2010overlapping}. IEFF integrates with these practices by using pre-rollout validation (QRT) and continuous monitoring to ensure safe feature rollouts.

\paragraph{Limitations and future work.}
IEFF is most effective when feature changes can be decomposed into gradual, bounded steps. Features with strong nonlinear interactions or high model sensitivity may still require explicit retraining, as small coverage changes can produce disproportionate effects. Additionally, the current fading rate selection relies on pre-rollout QRT validation; developing adaptive, data-driven fading rate policies remains an open direction.

Test-time scaling with feature fading. An interesting future direction is to leverage feature fading for test-time scaling. The fading magnitude naturally encodes a per-feature uncertainty signal, opening several avenues: (1) adaptive fading schedules that adjust decay rates based on real-time feature freshness, (2) ensembling predictions over multiple fading configurations for robustness, and (3) using fading magnitude as a confidence signal to trigger more expensive inference pathways only when predictions are unreliable. These approaches connect feature staleness handling with the broader test-time compute scaling paradigm.

\paragraph{Summary.}
Overall, prior work has addressed feature management, feature selection, retraining cost, and drift mitigation largely in isolation. IEFF fills an important gap by introducing a system-level mechanism for retrain-free, incremental feature rollouts in production ranking systems.

\section{Acknowledgements}
We thank the Fading V-team members and cross-functional partners for their close collaboration and contributions over the past several halves.

\bibliographystyle{ACM-Reference-Format}
\bibliography{paper}

\clearpage
\newpage

\end{document}